\documentclass[fleqn,twoside]{article}
\usepackage{espcrc2}
\usepackage{epsfig}

\newcommand{\AmS}{{\protect\the\textfont2
  A\kern-.1667em\lower.5ex\hbox{M}\kern-.125emS}}

\title{
\vskip -80pt
\mbox{} \hfill ITP Budapest 585\\
\vskip 25pt
Lattice study of the Coleman--Weinberg mass in the SU(2)-Higgs model
    \thanks{Presented by F. Csikor}}

\author{F. Csikor\address[ELTE]{Institute for Theoretical Physics,  
        E\"otv\"os University, \\
        P.O. Box 32, H-1117 Budapest, Hungary} 
        Z. Fodor\addressmark and
        P. Kov\'acs\addressmark }
\begin{document}

\begin{abstract}
Radiative symmetry breaking is a well known phenomenon in perturbation theory.
We study the problem in a non-perturbative framework, i.e. lattice
simulations. The example of the bosonic sector of the SU(2)-Higgs model is 
considered.  We determine the minimal scalar mass which turns 
out to be  higher than the mass value given by
1-loop continuum perturbation theory.
\vspace{1pc}
\end{abstract}

\maketitle

\section{INTRODUCTION}

Symmetry breaking is a fundamental concept in particle physics. 
The most popular mechanism is the Higgs mechanism. Also  radiative 
symmetry breaking (Coleman-Weinberg (CW) mechanism \cite{CW}) is well known.
The CW mass is obtained when the tree level mass parameter is zero, i.e. 
the Higgs mass is induced by radiative corrections. The value of the 
CW mass is near to  the smallest possible Higgs mass.

Connected to the CW mechanism is the problem of vacuum stability, 
i.e. the value of the minimal Higgs mass e.g. in the Standard Model.
Clearly the latter problem would be more interesting to study in a
non-perturbative framework. However, putting a chiral top quark on the 
lattice is difficult, therefore in this contribution 
we study symmetry breaking on the lattice only in the bosonic sector
of the SU(2)-Higgs model.  Still this is a difficult two scale problem, 
since the CW mass is an order of magnitude smaller than the W mass.

\section{LATTICE FORMULATION}
The lattice action  in standard notation is:
\begin{eqnarray}
&& S[U,\varphi]= 
 \beta \sum_{pl}
\left( 1 - {1 \over 2} {\rm Tr\,} U_{pl} \right) 
\nonumber \\
&&+ \sum_x \left\{ {1 \over 2}{\rm Tr\,}(\varphi_x^+\varphi_x)+
\lambda \left[ {1 \over 2}{\rm Tr\,}(\varphi_x^+\varphi_x) - 1 \right]^2
\right. \nonumber \\
&&\left.
-\kappa\sum_{\mu=1}^4
{\rm Tr\,}(\varphi^+_{x+\hat{\mu}}U_{x,\mu}\,\varphi_x) \right\}, 
\end{eqnarray}
where $U_{x,\mu}$ denotes the SU(2) gauge link variable,  $U_{pl}$ 
the path-ordered product of the four $U_{x,\mu}$ around a
 plaquette;
$\varphi_x$ stands for the Higgs field.
The continuum bare parameters are given:
$g^2=4/\beta$, $\lambda_c=\lambda/4/\kappa^2$, $\varphi_c=\sqrt{
2\kappa/a}\varphi
$ and $a^2 m_0^2=(1-2\, \lambda)/\kappa -8$. We choose $g^2=$0.5, which 
gives renormalized gauge couplings close to the Standard Model value.

The CW mechanism is an inherently perturbative notion. Therefore, on the 
lattice we want to determine the smallest Higgs mass possible (which is not
far from the CW mass).
The lattice action is bounded from below for positive $\lambda$,
while it is not for $\lambda  < 0$.
Thus $\lambda $=0 is the interesting point relevant to the
minimal and CW masses. Since the Higgs mass is a monotonous function of
$\lambda/\kappa^2$, the limit $\lambda  \rightarrow 0$
gives the smallest Higgs mass possible. 
Our aim is to determine the Higgs mass in this region and try to interpret
the results in terms of the CW mechanism.

\section{SIMULATIONS}

The lattice simulation goes in the standard way as described e.g. in 
\cite{CFHJM}.
It is essential to simulate along a line of constant physics (LCP).
For CW we define the LCP with 
$\lambda$=0, $g_R^2$=fixed ($g_R$ is the renormalized gauge coupling,
determined from a fit to the static potential) and the hopping parameter 
$\kappa$ tuned to the finite temperature phase transition point.

We simulate on $L_t$ $ \cdot$ $ L_s^3$ hot lattices  ($L_t  << L_s$) and
perform an $L_s \rightarrow \infty$ extrapolation.
Next we extrapolate to the continuum limit ($L_t \rightarrow \infty$).
The physical lattice sizes for the various $L_t$'s are kept constant by
choosing appropriate lattice extensions. The extrapolations to 
$L_s = \infty$ were carefully performed in order to eliminate finite size
effects.

In practice we take $\lambda$ very small ($5 \cdot 10^{-6}$,
$2.875 \cdot 10^{-5}$, $5.25 \cdot 10^{-5}$) and fix $\beta =8$. We 
found $\kappa_{cr}$ by {\em constrained simulation}, which is an  
appropriate method for strong phase transitions. 

The physical scale is found for each $\lambda$, $\beta$, $\kappa_{cr}$
  by performing simulations at $T$=0 and fixing $M_W$=80 GeV. The quantity
we extrapolate is $R_{HW}=M_H/M_W$.

Note that our LCP is not necessarily close to an LCP defined by the 
perturbative renormalization group running of $\lambda$ and $\beta$. 
In particular this is the case for the smallest $\lambda$ value given above.

\section{RESULTS}

\begin{figure}[t!]
\caption{Higgs mass squared as a function of $\lambda/\kappa_{cr}^2$.
The masses corresponding to the three highest values of
$\lambda/\kappa_{cr}^2$ are from \cite{CFHJM,FHJJM}, while the other point are
new simulations.}
\label{figure:1}
\epsfig{file=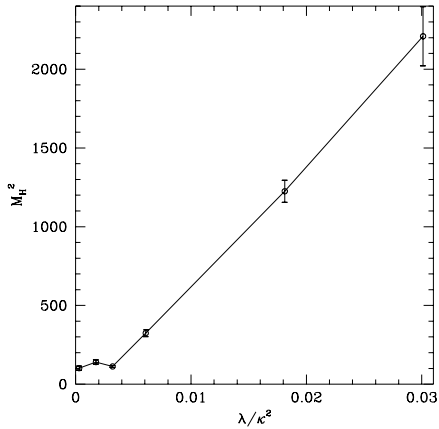,width=6.cm,angle=0} \nonumber
\end{figure}

In one loop perturbation theory one gets
\begin{equation}
M_{H}^2=v^2 \cdot (8\lambda_c+12 B),  \, \, \, \, B=\frac{9g^4}{1024\pi^2}
\end{equation}
The CW case corresponds to $\lambda_c=-B/2$. The minimal Higgs mass is for
$\lambda_c=-B$.

Our results for the Higgs mass squared as a function of $\lambda/\kappa_{cr}^2$ 
are given in  Fig. \ref{figure:1}. It is convenient to plot the 
Higgs mass squared as a function of $\lambda/\kappa_{cr}^2$, since (at tree
level) this is proportional to the continuuum quartic coupling. Similar plots 
for higher Higges masses have  been presented in \cite{A}. 
The straight line part of the curve on the right hand side of 
 the  figure obviously corresponds to the perturbative result. 
Even the slope is in reasonable aggreement with the perturbative prediction. 
Clearly, for small 
values of $\lambda/\kappa_{cr}^2$ the straight line has to flatten out, since 
otherwise we would obtain negative Higgs mass squares. This is what is seen 
on the figure. It is easy to extrapolate to $\lambda/\kappa_{cr}^2$=0. 
 We get $M_{H,min}^2=115.78 \pm 4.95$ $\rm GeV^2$ to be compared to the
perturbative value: 45.60 $\rm GeV^2$ (91.20 $\rm GeV^2$) corresponding to the
minimal Higgs mass (CW) cases. 
The lattice result is clearly much larger than 
the perturbative value. 

One would expect that the perturbative and
nonperturbative results are close to each other, since the quartic coupling is
small in this region. However, it is clear that the LCP choosen brings in a 
true nonperturbative feature in the  lattice  calculation. The CW LCP is
namely very far from the renormalization group LCP for small $\lambda$. 
Still, the large value of
the smallest nonperturbative Higgs mass is a surprising result. We have
carefully studied the finite size effects and believe that the discrepancy is
caused  most probably by the fact that our simulations are still far from the
continuum limit.

\section{ACKNOWLEDGEMENT}

This work was partially supported by Hungarian Science Foundation 
Grants No. OTKA-T34980/T29803/-M37071/OM-MU-708

\end{document}